\documentclass[pre,amsmath,showpacs]{revtex4}
\usepackage{graphicx} 
\usepackage{color}
\usepackage{bm}
\usepackage{epsfig}
\usepackage{latexsym}
\usepackage{pifont}
\begin{document}
\newcommand{\be}{\begin{equation}}
\newcommand{\ee}{\end{equation}}
\newcommand{\bea}{\begin{eqnarray}}
\newcommand{\eea}{\end{eqnarray}}

\def\Plus{\texttt{+}}

\title{Optimal search in $\bm{ E.coli}$ chemotaxis}
\author{Subrata Dev and Sakuntala Chatterjee}
\affiliation{Department of Theoretical Sciences, S. N. 
Bose National Centre for Basic Sciences, Block - JD, Sector - III, Salt Lake, 
Kolkata 700098, India }
\begin{abstract}

We study chemotaxis of a single {\sl E.coli} bacterium in a medium where the
nutrient chemical is also undergoing diffusion and its concentration has the
form of a Gaussian whose width increases with time. We measure
the average first passage time of the bacterium at a region of high nutrient
concentration. In the limit of very
slow nutrient diffusion, the bacterium effectively experiences a Gaussian
concentration profile with a fixed width. In this case we find 
that there exists an optimum width
of the Gaussian when the  average first passage time is minimum, {\sl i.e.},
the search process is most efficient. We verify the existence of the optimum
width for the deterministic initial position of the bacterium and also for
the stochastic initial position, drawn from uniform and steady state
distributions. Our numerical simulation in a model of a non-Markovian random
walker agrees well with our analytical calculations in a related
coarse-grained model. We also present our simulation results for the case 
when the  nutrient diffusion and bacterial
motion occur over comparable time-scales and the bacterium senses a time-varying
concentration field.  
\end{abstract}
\pacs{05.40.Jc, 05.10.Gg, 87.17.Jj}
\maketitle

%-------------------------------------------------------------
\section{Introduction} 

In a wide variety of physical systems, search process plays an important
role \cite{redner}.   
Examples can be found in systems such as animals searching for food
\cite{bell}, proteins searching for the binding site on DNA \cite{misha},
 or, diffusion-limited reactions \cite{raphael}. A search process
is often characterized by the first passage time, which is defined as the 
time taken to reach the target for the first time
 (or complete the search process). An efficient search process corresponds to a
short first passage time. Therefore it is crucial to determine how the first
passage time depends on the system parameters. The most efficient search
strategy is often determined by looking into the minimum of the first passage
time in this parameter space \cite{reset,sanjib}.

We consider the search process in one of the most well studied
biological systems, {\sl viz.}, {\sl E.coli}
 chemotaxis, which describes the motion
of {\sl E.coli} bacteria in response to a chemical concentration gradient
\cite{berg_book}. When such bacteria are placed in an inhomogeneous
concentration of a nutrient, they show a tendency to migrate towards the
nutrient-rich region \cite{adler,dahl}. 
We ask the questions: How long does it take for the bacteria to
find the most favorable region, and under what conditions is this search process
 most efficient?

\begin{figure}
\includegraphics[scale=0.5]{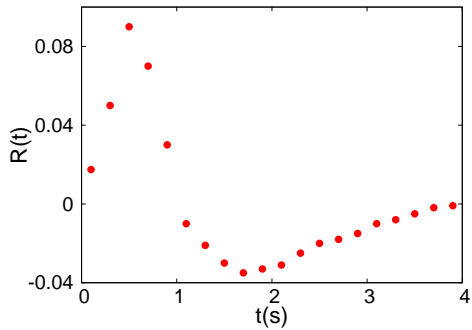}
\caption{(Color online) Bilobe response function of wild-type {\sl E. coli} used
 in our simulations. For computational
 simplicity, we have used a discrete sampling of the experimental data presented
in \cite{segall} instead of working with the complete data set. This did not
 affect our conclusions.}
\label{fig:bi}
\end{figure}

The motion of {\sl E.coli} takes place in two different modes: run and tumble.
During a run the bacteria move along a given direction with a fixed velocity
 $v \sim 20 \mu m /s$, and during a tumble they do not undergo appreciable
displacement but change their orientation randomly. In a homogeneous medium,
the average run duration is about $1s$, and at the end of one run the bacteria
go into the tumbling mode, which lasts for about $0.1s$ before another run in a
new direction starts \cite{brown,ryu}.
 In the presence of an inhomogeneous nutrient concentration
in the medium, the small size ($\sim 2 \mu m$) of an {\sl E.coli} cell  prevents
it from directly sensing the spatial gradient of the concentration field.
Therefore, to navigate to the nutrient-rich region, the bacteria rely
on temporal integration and modulate their run durations in different directions
in the medium via a memory kernel, shown in Fig. \ref{fig:bi}  \cite{berg}. 
 The kernel effectively compares the concentration experienced in the 
recent past to that in distant past, and if the difference is positive 
(negative) the run duration in the current direction is extended (shortened). 
Following the above description, the motion of a
single bacterium in a spatially varying chemical environment  
is modeled as a non-Markovian random walker with run and tumble modes, and the
switching rates between these modes depend on the 
nutrient concentration along its recent trajectory 
\cite{berg,gennes,clark,sc,celani}.

In many physical situations, the chemotaxis motion may take
place in a medium where the nutrient is also undergoing diffusion. Imagine a
situation where a puff of nutrient is injected into the medium such that
immediately after injection, all the chemical is concentrated in a narrow
region. As time goes on, this chemical spreads over the medium via diffusion
and at any given time its concentration profile has the form of a Gaussian. At
a certain stage of this nutrient diffusion, when the nutrient has already
spread through some distance in the medium, a bacterium is released somewhere
 in the medium which would perform chemotaxis. If
the time scale of nutrient diffusion is much longer than that of bacterial
motion, the bacterium would effectively
 experience a Gaussian concentration profile of
fixed width. When the nutrient diffusion and bacterial motion occur over 
comparable time scales, then the concentration sensed by the
bacterium will be time dependent and can be described as a Gaussian whose width
keeps increasing (and peak height keeps decreasing) with time. 
The region around the peak of the Gaussian profile, where the nutrient
concentration is highest, is therefore the most favorable region for the
bacterium. 

 In this paper, we study chemotaxis motion of a single bacterium 
in the
presence of a Gaussian concentration field of the nutrient, using the model
of a non-Markovian random walker described above. 
As considered in the experiments involving bacteria in a micro-fluidic channel
 or capillary assay \cite{bergturner, binz, mannik}, we consider the motion of
 the bacterium in one dimension. We measure the
average first passage time of the bacterium at the neighborhood of the
Gaussian peak using Monte Carlo simulation.  
 In the limit of slow nutrient diffusion, if the chemotaxis starts at a stage
when the width of the Gaussian profile is $\sigma$, then this width 
 does not change appreciably during the search process. The 
average first passage time in this case shows a minimum
as a function of $\sigma$. In other
words, there is an optimum value of $\sigma$ for which the search
process is most efficient. This finding is interesting since $\sigma$ is a
parameter that can be easily tuned in an experiment, and our study shows that
when $\sigma$ is set at a special value, the bacterium becomes the most
efficient searcher and is able to find its favorable region in the shortest
possible time. We also perform analytical calculations of
mean first passage time within a
coarse-grained model which allows an approximate Markovian description of the
bacterial motion \cite{sc,celani}. 
We find reasonably good agreement between our analytics and numerics. We 
consider a deterministic initial condition as well as a stochastic initial
condition, drawn from uniform and steady state distribution. In all the cases
our numerical simulation and analytical calculations show the existence of an
optimum width of the nutrient concentration profile when the search
 process is most efficient.

In the case when the time scale of nutrient diffusion is comparable to that of
bacterial motion, the bacterium experiences a time-varying 
concentration profile, a Gaussian whose width increases with time.
 The search process now crucially depends on the nutrient diffusivity, as
well as the extent
of spread of the nutrient in the medium at the onset of chemotaxis. Our
simulations show that if the chemotaxis starts at an early stage of nutrient
 diffusion, when the width of
the Gaussian is still small, then the mean first passage time shows a minimum
as a function of the nutrient diffusivity. But if the chemotaxis starts
at a late stage, when the nutrient has already spread considerably in the
medium and the width of the Gaussian profile is large, the mean
first passage time increases monotonically with nutrient diffusivity.

In the next section, we present our model and summarize earlier results. 
Sections $3$ and $4$ contain our analytical and numerical results,
respectively, for a Gaussian nutrient concentration. In section $5$ we present
our results for a time-varying concentration field.
The  conclusion is presented in section $6$.

%------------------------------------------------------------
\section{Description of the model and earlier results} 

\subsection{Simulation in a model of a non-Markovian random walker}

Following \cite{celani,kafri,sc}, we model the motion of a single
bacterium in one dimension as a non-Markovian random walker 
 whose dynamics is governed by runs
and tumbles. During a run the bacterium moves along one particular direction
with a fixed velocity. The duration of a run is a stochastic variable and
follows a Poissonian distribution with a mean of $1s$ in a homogeneous medium.  
At the end of a run, the bacterium goes into a tumbling mode, in which
 it rotates about itself 
in a random fashion, without much net displacement, before it starts running
again in a new direction. Because the
 average tumble duration is much smaller than that of
runs, tumbles are modeled as instantaneous events 
which allow the bacterium to change its direction between two successive runs.
In our present one-dimensional model, the probability that the run direction
is changed (reversed) is denoted as $q$. 
In the presence of a nutrient concentration gradient
in the medium, the tumbling rate depends on the recent history. 
The probability that a running bacterium tumbles
during a time interval $[t,t+dt]$ is then given by 
\be
\frac{dt}{\tau}\left (1-\int_{-\infty}^t dt' R(t-t') c(t') \right ),
\label{eq:tum}
\ee
where $\tau$ is the mean run duration in a homogeneous environment, $c(t')$ is
the concentration experienced at a past instant $t'<t$, and $R(t)$ is the
response kernel. $R(t)$ contains information about the signaling pathway
present inside the bacterial cell, and it was measured experimentally for 
wild-type 
bacteria in \cite{segall,berg}. $R(t)$ was shown to have a bilobe shape,
with a relatively sharp positive lobe at smaller $t$ and a shallow
 negative lobe at larger $t$ that vanishes for $t \gtrsim 4s$ (see Fig.
\ref{fig:bi}). The area under the positive lobe is roughly equal to the area
 under the
negative lobe, and this ensures that the response kernel merely measures the
concentration gradient and is insensitive to any overall change in the
background concentration level. Because of this property, it is called an
adaptive response kernel.

We are interested in the linear response regime, where the integral in Eq.
\ref{eq:tum} is much less than unity. Within this linear theory, we can
decompose the 
above bilobe response kernel into $\delta$-function response kernels of
suitably chosen amplitudes, and from a superposition of the solutions for
these $\delta$-function kernels, we can obtain the solution for the full bilobe
response. Hence we first consider $R(t) = \alpha
\delta (t-\Delta)$ and analyze this case in detail, where we keep terms only
up to first order in $\alpha$.  Later, we generalize our
results for the full response kernel.

\subsection{Analytical calculation in a coarse-grained Markovian description}

In an earlier study \cite{sc} a simple coarse-grained model was proposed for
describing a single bacterium in a concentration gradient in one dimension.
The bacterium was assumed to be confined in a one-dimensional box of length $L$
with reflecting boundary walls.  Although the dependence of the tumbling rate
on the past trajectory makes the underlying run-tumble motion 
non-Markovian, one can still expect that at a coarse-grained level, a
Markovian description might be possible. For this purpose, one can coarse-grain
over a time scale which is much larger than the typical run duration $\tau$
and assume that the average bacterial density within the spatial resolution of
coarse-graining $v \tau$ has a Markovian dynamics. For the time evolution
of this coarse-grained density $P(x,t)$  at point $x$, at time $t$, the 
following Fokker-Planck equation was formulated in \cite{sc}: 
\be
\partial_t P(x,t)=-\partial_x \left [V(x)P(x,t)-\partial_x (D(x)P(x,t)) \right
],
\label{eq:coarse}
\ee      
which is the equation for a random walker with position-dependent drift and
diffusion. In \cite{celani} the past memory of the cell trajectory was
included in the description by introducing additional `internal variables', and
the resulting process, characterized by a larger number
 of degrees of freedom, now becomes Markovian. Starting from this  Markovian
description, using the homogenization method \cite{erban,xue} the same
coarse-grained equation as above was obtained \cite{celani}.

The chemotactic
 drift velocity $V(x)$ and the diffusivity $D(x)$ in Eq. \ref{eq:coarse}
depend on the nutrient concentration profile $c(x)$, and the
dependence can be derived from the microscopic dynamics. In  an earlier
calculation by de Gennes an approximate expression for the drift velocity was
obtained  within the simplifying assumption that whenever a running bacterium 
tumbles, its past memory is lost. Considering the response function 
$R(t) = \alpha \delta (t-\Delta)$, the resulting expression 
\be
V(x) = \alpha\frac{v^2\tau}{2q} e^{-\frac{2 q \Delta}{\tau}} \partial_x c(x)  
\label{eq:vel}
\ee
was found to show good agreement with the simulation results \cite{sc}. To
calculate the diffusivity $D(x)$, the effective tumbling frequency was
calculated within the coarse-grained model by averaging over a population of
non-interacting bacteria within the coarse-graining length scale. Although the
tumbling frequency of a single bacterium depends on the details of its past
trajectory, this dependence is lost when averaged over a large number of
bacteria with all possible run directions. The average tumbling frequency
at a position $x$ can be shown to be $[1-\alpha c (x)]/\tau$, from
 which the diffusivity turns out to be \cite{sc} 
\be
D(x) = \frac{v^2 \tau}{2q}[1+\alpha c(x)].
\label{eq:diff}
\ee  
 
Using Eqs. \ref{eq:vel} and \ref{eq:diff}, it can be easily shown from Eq.
\ref{eq:coarse} that in the steady state the bacterial density $P(x)$ has the
form 
\be
P(x) = P_0 + \alpha P_0 (e^{-\frac{2q\Delta}{\tau}}-1) \left ( c(x) -
 P_0 \int_0^L c(x) dx \right ),
\label{eq:steady}
\ee
where $P_0=1/L$ and the last term takes care of the normalization.  

In the next section, we use the above coarse-grained model to calculate the
mean first passage time of the bacterium at the nutrient-rich region. Note that
in contrast to
 the steady state behavior, studied in \cite{sc}, we study first-passage
properties, away from steady state.

%---------------------------------------------------------------
\section{Analytical calculation of first passage time}

Let $P(y,t|x,0)$ be the conditional probability to find the bacterium at
position $y$ at time $t$, given that it started  at $x$ at $t=0$. This
conditional probability follows the backward Fokker-Planck equation: 
\be
\partial_t P(y,t|x,0) = V(x)\partial_x P(y,t|x,0) + D(x)\partial_x^2
P(y,t|x,0).
\label{eq:bwd}
\ee   
To find out the first passage time at a certain point $x_0$ (which we call the
target), we consider an
absorbing boundary condition at $x_0$, in addition to the reflecting boundary
at $x=0$ and $x=L$. Without any loss of generality, we perform all our
measurements for $x <x_0$. The survival probability $G(x,t)$ that starting
from an initial position $x$, the bacterium will not reach the target until 
time $t$ can be written as $G(x,t) = \int_0^{x_0} dy  P(y,t|x,0) $. From  
Eq. \ref{eq:bwd} it follows that $G(x,t)$ satisfies the following equation:
\be
\partial_t G(x,t) = V(x) \partial_x G(x,t) + D(x) \partial_x^2 G(x,t),
\ee
with the initial condition $G(x,0)=1$. The reflecting and absorbing boundary
conditions are implemented as $\partial_x G(0,t) =0$ and $G(x_0,t)=0$,
respectively.

By definition, $G(x,t)$ is the probability that the first passage time is
 larger than $t$, and hence the first passage time distribution is given
 by $-\partial_t G(x,t)$. Mean first passage time $T(x) =- \int_0^\infty
dt\; t \; \partial_t G(x,t) = \int_0 ^ \infty dt \; G(x,t) $, which follows the
equation
\be
V(x) \partial_x T(x) + D(x) \partial_x ^2 T(x) =-1.
\ee 
The solution of this equation has the form 
\be
T(x) = \int_x ^ {x_0} \frac{dy}{\psi(y)} \int_0 ^ y \frac{\psi(z)}{D(z)},
\ee  
where $\psi(x) = \exp \left [ \int_0 ^ x dx' V(x') / D(x') \right ] $.
Now, using Eqs. \ref{eq:vel} and \ref{eq:diff} and keeping terms only upto
first order in $\alpha$, one can write $\psi(x) = \exp \left [\alpha e^{-2 q
\Delta /\tau} \{c(x)-c(0)\}  \right ] = 1+\alpha e^{-2 q \Delta /\tau}
 [c(x)-c(0)] $. After a few steps of simple algebra we finally have 
\bea 
\nonumber
T(x) =& \frac{2q}{v^2 \tau} [ \frac{x_0^2 -x^2}{2} -\alpha(1-e^{-2q\Delta /
 \tau}) \int_x^{x_0} dy \int_0^y dz\; c(z)\\ 
& - \alpha e^{-2q\Delta / \tau} \int_x ^{x_0} dy\; y \; c(y)  ],
\label{eq:tx}
\eea

which can be written in the form $T(x) = T_0(x) + \alpha T_1(x)$, where
$T_0(x)$ stands for the mean first passage time for an ordinary Brownian motion
in the absence of any concentration gradient and $T_1(x)$ gives the first order
correction term when  position-dependent drift and diffusion are
present due to a spatially varying concentration field $c(x)$ \cite{angelani}.
 In the rest of this work we focus on  $T_1(x)$.

For a Gaussian concentration profile $c(x) = \dfrac{\exp [-(x-\overline{x})^2/2
\sigma^2]}{\sqrt{2 \pi \sigma^2}}$ the drift velocity $V(x)$ and
diffusivity $D(x)$ show rapid variation close to the peak at $\overline{x}$.
In our coarse-grained description, which allows for analytical treatment, we
deal with length scales much larger than the mean free path of the bacteria, and
any spatial variation that occurs over a smaller length scale must be
neglected in our coarse-grained model. When $\sigma$ is not too large, the
variation of $V(x)$ and $D(x)$ around the peak is too fast to be considered
in our coarse-grained formalism. Because of this, we choose the target position $x_0$ slightly away from the peak such that both the initial position $x$
 and the target lie on the same side of the peak. 
For our choice of $x<x_0<\overline{x}$ we use the Gaussian $c(x)$ in 
 Eq. \ref{eq:tx}  and get
\bea
\nonumber
T_1(x) &=& \frac{2q}{v^2 \tau} \Bigg[ \frac{1}{2} Erf 
\left (\frac{\overline{x} - x_0}{\sqrt{2} \sigma} \right ) 
\left (e^{-\frac{2q\Delta}{\tau}} (2 \overline{x} -x_0) -(\overline{x}-x_0)
\right ) \\
\nonumber
&&+ \frac{1}{2} Erf \left (\frac{\overline{x} - x}{\sqrt{2}
\sigma} \right ) \left ( e^{-\frac{2q\Delta}{\tau}} (x-2 \overline{x})-(x-
\overline{x}) \right )\\ \nonumber
&&+\frac{1}{2} Erf \left (\frac{\overline{x} }{\sqrt{2} \sigma} \right )
 \left ( \left ( e^{-\frac{2q\Delta}{\tau}} -1 \right )(x_0-x) \right )\\ 
 && +\left ( e^{ -\frac{(\overline{x}-x_0)^2}{2 \sigma^2}} -
 e^{ -\frac{(\overline{x}-x)^2}{2 \sigma^2}} \right ) 
\left ( 2e^{-\frac{2q\Delta}{\tau}} -1 \right )\frac{\sigma}{\sqrt{2 \pi}}
 \Bigg].
\label{eq:t1}
\eea

So far we have considered the first passage time with a deterministic initial
condition, where the bacterium always starts from a fixed initial position
$x$. Now we consider the case of stochastic initial position  
when $x$ can take any value within the interval 
$0 <x<x_0$; that is, the initial position can lie anywhere between the 
left boundary and the target, with a certain distribution function.
 We consider two specific cases: ($i$) when $x$ follows a uniform distribution
 $P_0$ and ($ii$) when $x$ is drawn from the steady
state distribution $P(x)$ in Eq. \ref{eq:steady}.  
 In the first case, the $\alpha$ order correction in the 
first passage time can be obtained by 
simply integrating Eq. \ref{eq:t1} over $x$:
\begin{eqnarray}
\nonumber
T_1^{(u)} &=& P_0\int_0^{x_0} dx T_1(x)\\
\nonumber
&=& \frac{2 q P_0}{ v^2 \tau} 
\Bigg[ \frac{1}{4} \left\{Erf(\frac{\bar{x}}{\sqrt{2}\sigma})-
Erf(\frac{\bar{x}-x_0}{\sqrt{2}\sigma})\right\} \\ \nonumber
&&\left\{e^{-\frac{2q\Delta}{\tau}}
(x_0^2-3\bar{x}^2-3\sigma^2)-(x_0^2-\bar{x}^2-\sigma^2)\right\}\\  
&&+\frac{1}{2\sqrt{2\pi}}\sigma(3e^{-\frac{2q\Delta}{\tau}}-1)
\left ( (x_0+\bar{x})  e^{-\frac{(x_0-\bar{x})^2}{2\sigma^2}} -\bar{x}
 e^{-\frac{\bar{x}^2}{2\sigma^2}} \right ) \Bigg]
\label{eq:uni}
\end{eqnarray} 
For case ($ii$) the initial position $x$ follows the steady state
distribution in Eq. \ref{eq:steady}.
The mean first passage time is then written as 
\bea
\int_0 ^{x_0} dx P(x) T(x) &=& \int_0 ^{x_0} dx \Bigg [ P_0 + 
\alpha P_0 (e^{-\frac{2q\Delta}{\tau}}-1)\\ \nonumber
&& \left (c(x) - P_0\int_0^L c(x) dx \right ) \Bigg ] \left [ T_0(x) + 
\alpha T_1 (x)\right ].
\eea  
For a Gaussian form of $c(x)$ the $\alpha$ order term becomes
\bea \nonumber
T_1^{(s)} &=& P_0\int_0^{x_0} dx T_1(x)
 + P_0 \frac{(e^{-\frac{2q\Delta}{\tau}}-1)}
{\sqrt{2 \pi \sigma^2}} \int_0^{x_0} dx 
 e^{-\frac{(x-\bar{x})^2}{2\sigma^2}}T_0(x) \\
&&-P_0^2(e^{-\frac{2q\Delta}{\tau}}-1) Erf \left (
 \frac{x_0}{\sqrt{2} \sigma} \right
)  \int_0^{x_0} dx T_0(x),
\eea
where $T_0(x)$ and $T_1(x)$ are defined in Eqs. \ref{eq:tx} and \ref{eq:t1}.
 After
straight forward algebra the $\alpha$ order term in first passage time with
the steady state initial condition becomes
\bea
\nonumber
T_1^{(s)} &=&
\frac{2qP_0}{v^2 \tau} \Bigg [ \frac{1}{2} \left \{ Erf \left
(\frac{\bar{x}}{\sqrt{2}\sigma}\right ) -Erf \left (
\frac{\bar{x}-x_0}{\sqrt{2}\sigma} \right ) \right \} \\ \nonumber
&& \left (
e^{-\frac{2q\Delta}{\tau_R}} (x_0^2 - 2 \bar{x}^2 - 2 \sigma^2) - (x_0^2
-\bar{x}^2 -\sigma^2 ) \right ) \\ \nonumber
&&+ \frac{\sigma  }{\sqrt{2 \pi}} \left ( 2 e^{-\frac{2q\Delta}{\tau_R}} -1 
\right ) \left \{ (x_0+\bar{x}) e^{-\frac{(\bar{x} -x_0)^2}{2\sigma^2}}
 -\bar{x} e^{-\frac{\bar{x}^2}{2\sigma^2}} \right \} \\
 && -\frac{P_0 x_0^3}{3} \left ( 2
e^{-\frac{2q\Delta}{\tau_R}} -1 \right )  Erf \left
(\frac{\bar{x}}{\sqrt{2}\sigma} \right ) \Bigg] 
\label{eq:st}
\eea
The results in Eqs \ref{eq:t1}, \ref{eq:uni} and \ref{eq:st} are for an
impulse response kernel $R(t)=\alpha \delta (t-\Delta)$, and these can be easily
generalized for any arbitrary response function. 
In the next section, we measure the first passage time in simulation and
compare the result with above analytical calculation.

%-------------------------------------------------------------
\section{Simulation results on first passage time }

We perform a Monte Carlo simulation on a one dimensional box of size $L$, with
reflecting boundary walls at the two ends. 
In a time interval $dt$ the bacterium moves by a distance $vdt$.
At the end of each time step, we compute the tumbling probability, as in
Eq. \ref{eq:tum}. For 
an impulse response function $R(t) = \alpha \delta ( t -\Delta)$, the tumbling
probability at time $t$ is given by $dt / \tau \left (1-\alpha c[x(t-\Delta)]
\right )$, where $x(t-\Delta)$ is the position of the bacterium at a time
$\Delta$ back in the past and $c[x(t-\Delta)]$ is the concentration experienced
by the bacterium at that past instant of time. At the end of one time step the
bacterium attempts to tumble with this probability. If the tumbling attempt is
unsuccessful, it continues to move in the same direction with same velocity
$v$, and if the tumbling attempt is successful, the bacterium changes its
direction (in this one-dimensional case, it reverses the sign of $v$) with
probability $q$. Starting from a given initial position $x$ at $t=0$, we
measure the first passage time at a position $x_0$ and average over different
trajectories. In order to avoid the region with rapid spatial
 variation of the concentration field, 
we consider a target which is one mean free path away from the
Gaussian peak in the same direction as the starting position of the bacterium:
$x_0=\overline{x} - v \tau $.

\begin{figure}
\includegraphics[scale=.7]{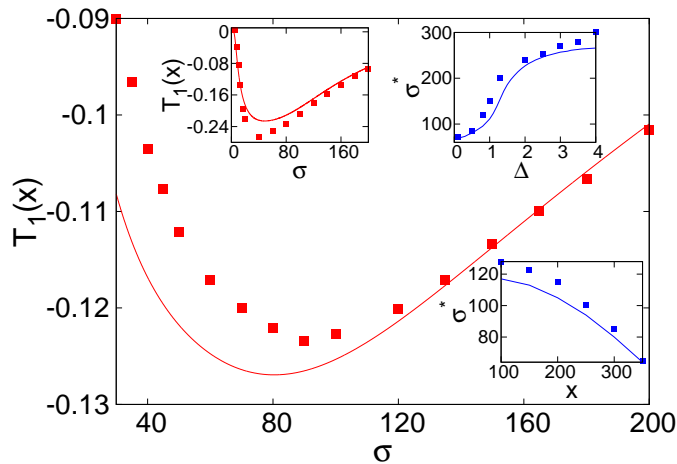}
\caption{(Color online) 
Results for the mean first passage time with a fixed initial
position. The main plot shows 
$T_{1}(x)$ (in seconds) as a function of standard deviation $ \sigma $ (in
$\mu m$) of the
Gaussian nutrient concentration field  with  $R(t)=\alpha\delta(t-\Delta)$ and
$\Delta=0.5s$, $x=300 \mu m$. These data have been averaged over $10^7$
different trajectories.
 The top right and bottom right insets show the variation of the
optimum width $\sigma^\ast$ (in $\mu m$) as a function of $\Delta$ (in
seconds) and the initial position 
$x$ (in $\mu m$),
 respectively. The error bar lies within $0.1 \mu m $. The top left
 inset shows $T_1(x)$ (in seconds) vs $\sigma$ (in $\mu m$) variation, averaged over $10^7$ trajectories, 
 for the bilobe response kernel, shown in Fig. \ref{fig:bi}. The discrete
symbols correspond to simulations and the solid lines correspond to
analytical calculations.  
Here $L=1000\mu m$, $x_{0}=490\mu m$,  $\bar{x}=500\mu m$, $q=0.5$, $\tau=1s$.} 
\label{fig:fix}
\end{figure}

In Fig. \ref{fig:fix} we show the variation of $T_1(x)$ with
 $\sigma$ (discrete symbols) and compare it with our analytical result in
 Eq. \ref{eq:t1} (solid
lines). We find reasonably good agreement between our simulation and
analytical calculation. For very small $\sigma$ the concentration
variation can be perceived only within a very narrow region around the peak
$\overline{x}$. Hence the bacterial trajectory starting from $x$ and ending
 at $x_0$, which does not cross the peak at $\overline{x}$, consists of
isotropic diffusion for the 
most part. As a result, in the limit of small $\sigma$
the mean first passage time is given by that for an ordinary Brownian motion
and is equal to $T_0(x)$ in Eq. \ref{eq:tx}, and the first order term $T_1(x)$
 goes
to zero. Similarly, in the limit of very large $\sigma$ the profile is almost
flat, and even in this case the motion is close to isotropic diffusion and
$T_1(x)$ vanishes. Our simulation and analytical calculation
are consistent with this simple argument.    

For intermediate $\sigma$ values, $T_1(x)$ must show a non-monotonic variation,
 since it vanishes for small and large $\sigma$. We find a minimum for $T_1(x)$
at a particular width $\sigma^\ast$. In other words, there exists an optimal
width $\sigma^\ast$ when the first passage time at the nutrient-rich region
becomes shortest and the bacterium becomes the most efficient searcher. 
The bottom right and top right 
insets in Fig. \ref{fig:fix} show the variation of the optimal width
$\sigma^\ast$ as a function of the initial position $x$ and the
 memory $\Delta$ of the bacterium, respectively. Note that
even in the Markovian limit, when the bacterium does not have any memory, and
 does not accumulate in the nutrient-rich region in the long time limit
\cite{sc}, its first passage properties still show the existence of an optimal
width when the search is most efficient.

For wild-type bacteria, the response kernel has a bilobe shape, and we can
reconstruct the kernel as a linear superposition of impulse response functions
with suitable amplitudes, as shown in Fig. \ref{fig:bi}. We use this
response kernel to calculate the mean first passage time for various $\sigma$. 
Even for this adaptive bilobe kernel, we find there exists an optimal width
when the mean first passage time hits a minimum. Our analytical calculations
show similar results. Interestingly, the value of the optimum width
$\sigma^\ast$ does not change even when the initial position and the target
position are varied (data not shown here). In other words, a wild-type {\sl
E.coli}  
bacterium becomes the most efficient searcher when placed in an environment
with a  
Gaussian concentration profile of a nutrient with a width of $\sim 50 \mu m$.

\begin{figure}
\includegraphics[scale=0.7]{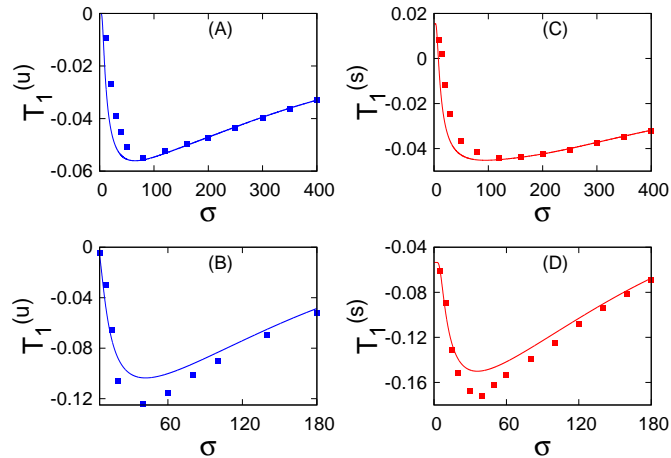}
\caption{ (Color online) Mean first passage time with stochastic 
initial positions. The top row shows data for impulse response
  $R(t)=\alpha\delta(t-\Delta)$, and the bottom row shows data 
for the bilobe response. 
(A) and (B) show data for a uniform initial condition, when $x$ can take any
value in the range $[0,x_0]$ with uniform probability $P_0$.
 (C) and (D) show data for a steady state initial condition, when the value of
$x$ in the range $[0,x_0]$ is drawn from the steady state distribution $P(x)$
in Eq. \ref{eq:steady}. Here the first passage time is measured in the units of
seconds, and $\sigma$ is in $\mu m$. The other simulation parameters are the
 same as in the main plot of Fig. \ref{fig:fix}. 
These data have been averaged over $10^7$ trajectories. 
The discrete symbols are for numerical data,
and the solid lines are for analytical calculations. }
\label{fig:unist}
\end{figure}

Instead of starting from a fixed position, when the initial position is a
random variable which can choose any value between the left boundary wall at
$x=0$  and the target at $x=x_0$ with a certain distribution, our results 
show the existence of an optimum
$\sigma$ that minimizes the first passage time. We have considered
initial positions chosen from uniform distribution as well as from steady state
distribution in Eq. \ref{eq:steady}.
 We present our data in Fig. \ref{fig:unist}. 
Note that for the choice of a steady state initial condition, 
$T_1^{(s)}$ does not vanish in the
limit of small $\sigma$ but approaches a constant value. In other words, even
when the width of the nutrient concentration profile
 is vanishingly small, the first
passage time of the bacterium is not the 
same as in a homogeneous medium. In fact, when the system is in the
 steady state, the bacterium has explored the full system
and has already experienced the narrow concentration profile present in the
middle of the box. The steady state measure $P(x)$ is therefore not the same as
$P_0$  but contains information about the narrow concentration field. This
gives rise to a non-vanishing $\alpha$ order correction term in the limit
$\sigma \rightarrow 0$.      

%--------------------------------------------------------------
\section{Time-varying concentration field}

In this section, we consider the case when the nutrient diffusion in the
medium occurs over a time scale comparable to that of bacterial motion. The
bacterium will then experience a time-varying concentration field. Our
 analytical formalism in section $3$ 
does not work in this case, and we study the system using numerical
simulations. The simplest
description of the nutrient concentration profile can be given by a Gaussian
whose width is increasing with time: $c(x,t) = \exp \left
( -\dfrac{(x-\overline{x})^2}{2(\sigma_0^2 + 2{\cal D}t) } \right )/ \sqrt{2 \pi (
\sigma_0^2 + 2{\cal D}t)}$, where $\sigma_0$ is the width at the time when the
chemotaxis motion starts, and ${\cal D}$ is the nutrient diffusivity.

\begin{figure}
\includegraphics[scale=0.7]{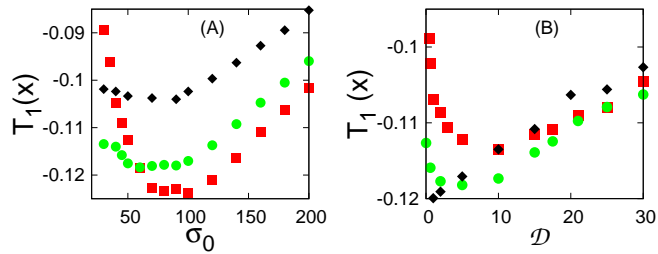}
\caption{(Color online) 
First passage time for time-dependent concentration of the nutrient.
(A) shows the variation of  $T_{1}(x)$ (in seconds) as a function of $
\sigma_0 $ (in $\mu m$) 
with $\cal D$ held fixed at $0.01 \mu m ^2 /s$ (red squares), $10 \mu m ^2
/s$ (green circles), and $37 \mu m ^2 /s$ (black diamonds). (B) shows
$T_{1}(x)$ (in seconds) vs $\cal D$ (in $\mu m ^2 /s$)
 plot for $\sigma_0 = 30 \mu m$ (red squares), $50 \mu
m$ (green circles) and $120 \mu m$ (black diamonds). The other simulation
parameters are the same as in Fig. \ref{fig:fix} main plot. These data are averaged over $10^8$ trajectories.}
\label{fig:cxt}
\end{figure}

The bacterial motion will depend on $\sigma_0$ and $\cal D$,
depending on the time scale $t_c \sim \sigma_0^2/{\cal D} $. For $t \ll t_c$ the
motion depends on $\sigma_0$, and for $t \gg t_c$ the motion  is
mainly controlled by $\cal D$. In the limit of very small $\cal D$, therefore,
one would expect the first passage time to be a function of $\sigma_0$ alone. 
In fact this is the limit when the nutrient diffusion is very slow, and during
the time interval of the 
first passage at the target, the width of  $c(x,t)$ changes very
little. In this limit, therefore, one expects results similar to those
in a static
concentration profile. Our simulation data in Fig. \ref{fig:cxt}A indeed
show that for small $\cal D$ there is an optimum width $\sigma_0$
 where $T_1(x)$ becomes 
minimum. As $\cal D$ increases, $t_c$ becomes smaller when $T_1(x)$
 does not show
much variation with $\sigma_0$ and the minimum becomes less and less
pronounced. In Fig. \ref{fig:cxt}A we verify this. 

In Fig. \ref{fig:cxt}B we show the variation of $T_1(x)$ against $\cal D$ 
for fixed
$\sigma_0$ values. For very large $\cal D$ the Gaussian profile
quickly flattens out, and the bacterial motion becomes an isotropic diffusion.
In this limit $T_1(x)$ becomes zero. For very small $\cal D$ values, the limit
for a static Gaussian profile is recovered, and (as shown in our data in Fig.
\ref{fig:fix}, main plot) $T_1(x)$ has 
a negative value that depends on $\sigma_0$.  
Therefore, for a given $\sigma_0$, as $\cal D$ is varied, $T_1(x)$
starts from a negative value at small $\cal D$ and  becomes
zero at large $\cal D$. Whether this variation is monotonic or not depends on
the choice of $\sigma_0$. Our data in Fig. \ref{fig:cxt}B show that for large
$\sigma_0$ the variation is monotonic but for small $\sigma_0$
a minimum is reached at a particular $\cal D$; that is, there is an optimum
diffusivity of the nutrient when the search is most efficient. For our various
choice of $\sigma_0$ values over a wide range 
(full data set not presented here), we
also notice that an optimum diffusivity is observed 
 whenever $\sigma_0$ is fixed at a value smaller than 
$\sigma^\ast$, the optimum width for the static 
concentration profile (see Fig. \ref{fig:fix} main plot). For $\sigma_0 >
\sigma ^\ast$, on the other hand, $T_1(x)$ increases monotonically with $\cal
D$.

The above observation tentatively indicates that it may be possible to describe
the results for the time-dependent nutrient concentration in terms of a static
concentration profile with an `effective width' $\sigma_e$. 
For a given value of $\sigma_0$ and $\cal D$ the width of
$c(x,t)$ keeps increasing during bacterial motion: at the start of the motion
the width is $\sigma_0$, and at the end of the first passage the average width
is $\sqrt{\sigma_0^2+2 {\cal D} T(x)}$. Let us assume that $\sigma_e$ is some
measure of the average or effective width experienced by the bacterium during
this process. Obviously, $\sigma_e$ is a function of both $\sigma_0$ and 
$\cal D$: for a
fixed $\sigma_0$, as $\cal D$ is varied, $\sigma_e \approx \sigma_0$ for very
small $\cal D$, and as $\cal D$ becomes very large, so does $\sigma_e$. In the
course of this variation, if $\sigma_e$ crosses $\sigma^\ast$, then $T_1(x)$
shows a minimum and if $\sigma_0 > \sigma^\ast$ such that $\sigma_e$ never
reaches $\sigma^\ast$ (because $\sigma_e$ can never fall below $\sigma_0$),
 then $T_1(x)$ shows a monotonic increase with $\cal D$. The above picture
explains our numerical data well. However, we would like to mention that we do
not yet have any mathematical expression for $\sigma_e$ in terms of $\sigma_0$
and $\cal D$. It would be interesting to directly verify the mechanism
proposed above. 

%------------------------------------------------------------------------

\section{Conclusion}

In this paper, we have considered the chemotaxis motion of a bacterium in a
medium where the nutrient is also undergoing diffusion and its concentration
profile is given by a Gaussian whose width increases with time. We
have measured the mean first passage time of the bacterium at the
neighborhood of the Gaussian peak. In the limit when the nutrient
diffusion is slow compared to the bacterial motion, the bacterium 
experiences an effectively static concentration profile, a Gaussian with a
fixed width, and in this regime we
calculate the mean first passage time analytically, within a
coarse-grained formalism. We find that the mean first passage time shows a
minimum as a function of the width of the Gaussian, which means that the
search process becomes most efficient at a certain optimum width. Our
numerical simulation matches the analytical result well.

For a time-dependent concentration profile, {\sl i.e.}, in the regime in which
 the
nutrient diffusion occurs over a time-scale comparable to bacterial motion, we
find that the first passage time is a function of nutrient diffusivity $\cal
D$ and the width $\sigma_0$ of the Gaussian at the onset of chemotaxis motion.
When $\cal D$ is held fixed at a small value, the mean first passage time
shows a minimum against variation of $\sigma_0$, as in the static case.
 But for large $\cal D$ the minimum becomes less pronounced. As a function of
$\cal D$, the mean first passage time shows a minimum if $\sigma_0$ is held
fixed at a small value. But no such minimum is observed when $\sigma_0$ is set
at a large value.

The existence of an optimal width of the nutrient concentration profile 
that makes the search process most efficient is
an interesting result. Apart from the mean first passage time, we have also
examined our conclusion by measuring the most probable first passage time.
Note that for a general process described by an equation of the form given in
Eq. \ref{eq:coarse}, the probability distribution for first passage time shows
a long tail which makes the mean first passage time much larger than the
typical (or most probable) one. Our numerical simulations (data not shown
here) show that the typical first passage time also becomes minimum at a
particular width whose value is very close to the one for which the mean
showed a minimum.

In \cite{celani} the chemotactic efficiency was characterized by a quantity
called `uptake' (defined as the total amount of nutrient encountered by 
the bacterium upto a certain time). It was shown that under
harsh environmental conditions, when the nutrient is scarce, the long time 
uptake can be
maximized for a particular shape of the response kernel. In contrast, we
consider a given response kernel and vary the parameter(s) characterizing the
concentration gradient of the nutrient in the medium to find the minimum first
passage time. In other words, for a given response kernel we find the optimum
environmental conditions for the fastest search process. It is easy to argue
that the long time uptake does not show any maximum in our case, when the
response kernel is fixed and the environmental conditions are varied. The
long time uptake decreases monotonically as the concentration gradient is
decreased.

Finally, it would be interesting to verify our conclusions in experiments. 
Recently, {\sl E.coli} chemotaxis has been studied in a micro-fluidic channel 
whose width is comparable to the bacterial mean free path \cite{bergturner,
binz, mannik}. In such a setup, the motion of the bacterium can be considered
to be effectively one-dimensional. It is possible to generate a  
Gaussian chemical concentration profile using techniques of diffusive
microfluidics \cite{kim}.
 The motion of the bacterium can be tracked to measure its first
passage properties. Our model predicts that for a static Gaussian profile of 
width $\sigma \sim 50 \mu m$, wild-type {\sl E.coli}
 have shortest first passage time.
However, it can be experimentally challenging to verify our results for a
time-dependent nutrient concentration profile. Most common chemoattractants
such as aspartate and serine
have diffusivity ${\cal D} \sim 1000 \mu m^2/s$  which is much larger than
 bacterial diffusivity. As a result, the chemical diffuses very quickly in the
medium, and initially localized concentration quickly flattens out. Thus
the bacterium experiences a very weak concentration gradient 
 and the chemotactic correction $T_1$ to its first passage time may
become too small for experimental detection.

\section{Acknowledgment}
 We acknowledge the computational facility provided through
 Thematic Unit of Execellence on Computational Material Science, funded by
Nanomission, Department of Science and Technology, India. 

%-------------------------------------------------------------

\end{document}